\documentclass[12pt,a4paper]{article}
\usepackage{epsfig,float,wrapfig}
\usepackage{amsmath}
\usepackage{amssymb}
\usepackage{bbm}
\newcommand{\be}{\begin{equation}}
\newcommand{\ba}{\begin{eqnarray}}
\newcommand{\ea}{\end{eqnarray}}
\def\a{\alpha}

\def\d{\delta}
\def\e{\epsilon}

\def\g{\gamma}

\def\j{\psi}

\def\l{\lambda}
\def\m{\mu}
\def\n{\nu}

\def\q{\theta}
\def\r{\rho}
\def\s{\sigma}

\def\P{\Pi}

\def\S{\Sigma}

\def\cs{{\cal S}}

\newcommand{\ti}{\tilde}

\newcommand{\pa}{\partial}

\newcommand{\im}{{\rm{i}}}

\newcommand{\nn}{\nonumber}
\newcommand{\tra}{{\rm{Tr}}}

\renewcommand{\title}[1]{\null\vspace{25mm}\noindent{\Large{\bf #1}}\vspace{10mm}}
\newcommand{\authors}[1]{\noindent{\large #1}\vspace{20mm}}
\newcommand{\address}[1]{{\center{\noindent #1\vspace{0mm}}}}
\renewcommand{\abstract}[1]{\vspace{17mm}
\noindent{\small{\em Abstract.} #1}\vspace{2mm}}     
\newcommand{\cbh}{\Hat{\Bar{c}}}

\newcommand{\dtt}{\Tilde{\Tilde{\partial}}}
\newcommand{\psibh}{\Hat{\Bar{\psi}}}
\DeclareMathAlphabet\mathbb  {U}{msb}{m}{n}
\DeclareFontFamily{U}{msb}{} \DeclareFontShape{U}{msb}{m}{n}{
  <5> <6> <7> <8> <9> gen * msbm
  <10> <10.95> <12> <14.4> <17.28> <20.74> <24.88> msbm10
  }{}

\makeatletter
\def\section{\@startsection{section}{1}{\z@}{-3.25ex plus -1ex minus
             -.2ex}{1.5ex plus .2ex}{\normalfont\bfseries}}
\def\subsection{\@startsection{subsection}{1}{\z@}{-3.25ex plus -1ex
                minus -.2ex}{1.5ex plus .2ex}{\normalfont\itshape}}

\renewenvironment{thebibliography}[1]
         {\section*{References}\frenchspacing\small
          \begin{list}{[\arabic{enumi}]}
         {\usecounter{enumi}\parsep=2pt\topsep 0pt
         \settowidth{\labelwidth}{[#1]}
         \leftmargin=\labelwidth\advance\leftmargin\labelsep
         \rightmargin=0pt\itemsep=0pt\sloppy}}{\end{list}}

\makeatother
\begin{document}   \setcounter{table}{0}
 
\begin{titlepage}
\begin{center}
\hspace*{\fill}{{\normalsize \begin{tabular}{l}
                              {\sf hep-th/0102103}\\
                              {\sf REF. TUW 01-03}\\
                              {\sf REF. UWThPh-2001-9}
                              \end{tabular}   }}

\title{\vspace{5mm} Deformed QED via Seiberg-Witten Map}

\vspace{20mm}

\authors {  \Large{A. A. Bichl$^{1}$, J. M. Grimstrup$^{2}$, L. Popp$^{3}$, M. Schweda$^{4}$,\\ R. Wulkenhaar$^{5}$ }}    \vspace{-20mm}

\vspace{10mm}
       
\address{$^{1,2,3,4}$  Institut f\"ur Theoretische Physik, Technische Universit\"at Wien\\
      Wiedner Hauptstra\ss e 8-10, A-1040 Wien, Austria}
\address{$^{5}$  Institut f\"ur Theoretische Physik, Universit\"at Wien\\Boltzmanngasse 5, A-1090 Wien, Austria   }

\footnotetext[1]{Work supported in part by ``Fonds zur F\"orderung der Wissenschaftlichen Forschung'' (FWF) under contract P14639-TPH.}

\footnotetext[2]{Work supported by The Danish Research Agency.}

\footnotetext[3]{Work supported in part by ``Fonds zur F\"orderung der Wissenschaftlichen Forschung'' (FWF) under contract P13125-PHY.}

\footnotetext[5]{Marie-Curie Fellow.}       
\end{center} 
\thispagestyle{empty}
\begin{center}
\begin{minipage}{12cm}

\vspace{10mm}

{\it Abstract.}   With the help of the Seiberg-Witten map for photons and fermions we define a $\q$-deformed QED at the classical level. Two possibilities of gauge-fixing are discussed. A possible non-Abelian extension for a pure $\q$-deformed Yang-Mills theory is also presented.

\end{minipage}\end{center}
\end{titlepage}

\section{Introduction}
The Seiberg-Witten map was originally discussed in the context of string theory, where it emerged from a 2D-$\s$-model regularized in different ways \cite{Seiberg:1999vs}. It was argued by Seiberg and Witten that the ordinary gauge theory should be gauge-equivalent to a noncommutative counterpart. The Seiberg-Witten map may be also introduced with the help of covariant coordinates. In this way the deformed gauge theory emerges as a gauge theory of a certain noncommutative algebra \cite{Madore:2000en}, \cite{Filk:1996dm} leading to the same results as in ref.\cite{Seiberg:1999vs}. 

In this note we apply the idea of the Seiberg-Witten map in order to define a $\q$-deformed QED at the classical level, including the corresponding Seiberg-Witten map for the infinitesimal local gauge transformation of the fermions. Using the definitions of the star-product \cite{Filk:1996dm} and the Seiberg-Witten map for the Abelian gauge field and the fermions we define a $\q$-deformed QED in terms of the deformation parameter $\q$ of a noncommuting flat space-time. This deformation parameter, which is treated as a constant external field with canonical dimension $-2$, leads to a field theory with infinitely many $\q$-dependent interaction vertices.

In order to prepare the quantization of this deformed QED (and possible non-Abelian extensions) we discuss the gauge-fixing procedure of the photon sector in the Seiberg-Witten framework. It is interesting to note that there are two possibilities of gauge-fixing. The linear one corresponds to the usual gauge-fixing of the undeformed Abelian model, whereas the second possibility is induced by a nonlinear gauge-fixing emerging from the noncommutative structure of the deformed QED. It is important to observe that the $\q$-deformed QED is still characterized by an Abelian gauge symmetry. The symmetry of the model is described by a linear BRST-identity.

The letter is organized as follows. Section 2 is devoted to a definition of the Seiberg-Witten map for photons and fermions. With these $\q$-expansions for the basic field variables of the model and the usual definition of the star-product \cite{Filk:1996dm} an Abelian gauge invariant deformed QED action is defined. The gauge-fixing procedure \`a la BRST \cite{Becchi:1976nq} will be discussed in section 3. Higher derivative gauge invariant monomials which are allowed due to the presence of $\q$ are analyzed in section 4. In fact such terms are required if one studies radiative corrections \cite{Bichl:2001nf} .

In a last step we present also a non-Abelian extension of the photon sector, the pure $\q$-deformed noncommutative Yang-Mills (NCYM) theory, involving the discussion of the invariant action, the gauge-fixing procedure and the higher derivative gauge invariant extensions of the action.

\section{Deformed QED via Seiberg-Witten Map}
The starting point is the undeformed QED with its local gauge invariance
\begin{equation}
S_{\rm inv} = \int d^4x\, \left[ \bar{\j} \left( \im \g^{\m}D_{\m}-m \right) \j - \frac{1}{4}\,F_{\m\n}F^{\m\n} \right], \label{Sinv}
\end{equation}
where
\begin{align}
D_{\m}\j &= \left( \pa_{\m}-\im A_{\m} \right) \j,
\\
F_{\m\n} &= \pa_{\m}A_{\n} - \pa_{\n}A_{\m}. \label{F}
\end{align}
$S_{\rm inv}$ is invariant with respect to infinitesimal gauge transformations:
\begin{align}
\d_{\l}A_{\m} &= \pa_{\m}\l, \nn
\\
\d_{\l}\j &= \im\l\j, \nn
\\
\d_{\l}\bar{\j} &= -\im \bar{\j}\l. \label{infgt}
\end{align}
In order to get a $\q$-deformed QED in the sense of Seiberg and Witten \cite{Seiberg:1999vs} one defines
\begin{align}
\hat{A}_{\m} &= A_{\m} + A'_{\m} \left( A_{\n};\q^{\r\s} \right) + O(\q^{2}), \nn
\\
\hat{\j} &= \j + \j' \left( \j,A_{\n};\q^{\r\s} \right) + O(\q^{2}), \nn
\\
\hat{\l} &= \l + \l' \left( \l,A_{\n};\q^{\r\s} \right) + O(\q^{2}) \label{SWQED}, 
\end{align}
with the corresponding Seiberg-Witten maps given by 
\begin{align}
\hat{A}_{\m} \left( A_{\n} \right) + \hat{\d}_{\hat{\l}}\hat{A}_{\m}(A_{\n}) &= \hat{A}_{\m} \left( A_{\n} + \d_{\l}A_{\n} \right), \label{SW}
\\
\hat{\j} \left( \j,A_{\n} \right) + \hat{\d}_{\hat{\l}}\hat{\j}(\j,A_{\n}) &= \hat{\j} \left(\j+\d_{\l}\j,A_{\n} + \d_{\l}A_{\n} \right).
\end{align}
The deformed infinitesimal gauge transformations are defined as usual \cite{Hayakawa:1999zf}:
\begin{align}
\hat{\d}_{\hat{\l}}\hat{A}_{\m} &= \pa_{\m}\hat{\l} + \im \left[ \hat{\l},\hat{A}_{\m} \right]_{\star} = \pa_{\m}\hat{\l} + \im\hat{\l}\star\hat{A}_{\m} - \im\hat{A}_{\m}\star\hat{\l}, \nn
\\
\hat{\d}_{\hat{\l}}\hat{\j} &= \im \hat{\l}\star\hat{\j}, \nn 
\\
\hat{\d}_{\hat{\l}}\psibh &= - \im \psibh \star \hat{\l}, \label{gt}
\end{align}
where we have used the star-product \cite{Filk:1996dm}. Expanding (\ref{gt}) in powers of $\q$ one gets
\begin{align}
\hat{\d}_{\hat{\l}}\hat{A}_{\m} &= \pa_{\m}\hat{\l} - \q^{\r\s}\,\pa_{\r}\l\,\pa_{\s}A_{\m} + O(\q^2), \nn
\\
\hat{\d}_{\hat{\l}}\hat{\j} &= \im \hat{\l}\hat{\j} - \frac{1}{2}\,\q^{\r\s}\,\pa_{\r}\l\,\pa_{\s}\j + O(\q^2). \label{nA}
\end{align}
With (\ref{SW})--(\ref{gt}) one gets the following solutions of (\ref{SWQED}):
\begin{align}
A'_{\m} &= -\frac{1}{2}\,\q^{\r\s}\,A_{\r} \left( \pa_{\s}A_{\m} + F_{\s\m} \right), \nn
\\
\j' &= -\frac{1}{2}\,\q^{\r\s}\,A_{\r}\,\pa_{\s}\j, \nn
\\
\l' &=- \frac{1}{2}\,\q^{\r\s}\,A_{\r}\,\pa_{\s}\l. \label{'}
\end{align}
In terms of the commutative fields  (\ref{F}) the deformed field strength becomes
\begin{align}
\hat{F}_{\m\n} &= \pa_{\m}\hat{A}_{\n} - \pa_{\n}\hat{A}_{\m} + \q^{\r\s}\,\pa_{\r}A_{\m}\,\pa_{\s}A_{\n} + O(\q^2) \nn
\\
&= F_{\m\n} + \q^{\r\s} \left( F_{\m\r}\,F_{\n\s} - A_{\r}\,\pa_{\s}F_{\m\n} \right) + O(\q^2).
\end{align}
The infinitesimal gauge transformation of $\hat{F}_{\m\n}$ is calculated with $\d_{\l}A_{\m}=\pa_{\m}\l$ (from now on we omit the symbol $O(\q^2)$)
\begin{align}
\hat{\d}_{\hat{\l}}\hat{F}_{\m\n} &= - \q^{\r\s}\,\pa_{\r}\hat{\l}\,\pa_{\s}\hat{F}_{\m\n} \nn
\\
&= - \q^{\r\s}\,\pa_{\r}\l\,\pa_{\s}F_{\m\n}.
\end{align}
Now we are able to write down the analogue of (\ref{Sinv}), leading to the following generalization
\begin{align}
\hat{\S}_{\rm inv} &= \int d^4x\, \left[ \psibh \star \left( \im \g^{\m}\hat{D}_{\m}-m \right) \hat{\j} - \frac{1}{4}\,\hat{F}_{\m\n}\star\hat{F}^{\m\n} \right] \nn
\\
&= \int d^4x\, \left[ \psibh \left( \im \g^{\m}\hat{D}_{\m}-m \right) \hat{\j} - \frac{1}{4}\,\hat{F}_{\m\n}\hat{F}^{\m\n} \right],
\end{align}
with
\begin{align}
\hat{D}_{\m}\hat{\j} = \pa_{\m}\hat{\j} -\im \hat{A}_{\m} \star \hat{\j}. \nn
\end{align}
Using the above expansions in $\q$ one gets
\begin{multline} \label{HSinv} 
  \S_{\rm inv} = \int d^4x\, \left\{ \bar{\j}\,\im \g^{\m} \left[ D_{\m}     \j - \frac{1}{2}\,\q^{\r\s} \left( \pa_{\m}A_{\r}\,\pa_{\s}\j + A_{\r}\,         \pa_{\m}\pa_{\s}\j \right) \right. \right.  \\
  \left.+ \frac{\im}{2}\,\q^{\r\s} \left( A_{\r} \left( \pa_{\s}A_{\m} + F_{\s     \m} \right) \j + A_{\m}A_{\r}\,\pa_{\s}\j - \,\im\, \pa_{\r} A_{\m} \pa_{\s}       \j\right) \right]  \\
  - \frac{1}{2}\,\q^{\r\s}\,A_{\r}\,\pa_{\s}\bar{\j}\,\im\,\g^{\m}D_{\m}\j - m     \bar{\j}\j +\frac{1}{2} m \q^{\r\s}\pa_{\r}A_{\s}\bar{\j}\j  \\
  \left. - \frac{1}{4}\,F_{\m\n}F^{\m\n} - \frac{1}{2}\,\q^{\r\s} \left( F_{\m     \r}F_{\n    \s}F^{\m\n} - \frac{1}{4}\,F_{\r\s}F_{\m\n}F^{\m\n} \right)          \right\}. 
\end{multline}
Equation (\ref{HSinv}) contains non-renormalizable vertices of dimension $6$. The quantity $\q^{\r\s}$ will be considered as a classical unquantized external field with dimension $-2$. The $\q$-deformed action (\ref{HSinv}) is invariant with respect to (\ref{infgt}).
\section{Gauge-Fixing \`{a} la BRST}
In order to quantize the system (\ref{HSinv}) we need a gauge-fixing term which allows of the calculation of the photon propagator. This can be done in a twofold manner: Corresponding to the usual BRST-quantization procedure one replaces the infinitesimal gauge parameters $\l$ and $\hat{\l}$ by Faddeev-Popov ghost fields $c$ and $\hat{c}$ leading to the following BRST-transformations for the Abelian structure
\begin{equation}
sA_{\m} = \pa_{\m}c, \quad  s\j = \im c \j.\label{sA}
\end{equation}
Corresponding to the noncommutative structure of (\ref{nA}) one gets
\begin{align}
\hat{s}\hat{A}_{\m} &= \pa_{\m}\hat{c} - \q^{\r\s}\,\pa_{\r}c\,\pa_{\s}A_{\m}, \nn
\\
\hat{s}\hat{\j} &= \im \,\hat{c}\,\hat{\j} - \frac{1}{2}\,\q^{\r\s}\,\pa_{\r}c\,\pa_{\s}\j. \label{hsA}
\end{align}
Nilpotency of (\ref{sA}) and (\ref{hsA}) implies
\begin{align}
sc &= 0, \label{sc}
\\
\hat{s}\hat{c} &= -\frac{1}{2}\,\q^{\r\s}\pa_{\r}c\,\pa_{\s}c.
\end{align}
Clearly, due to (\ref{'}) one has also
\begin{align}
\hat{c} = c - \frac{1}{2}\,\q^{\r\s}A_{\r}\,\pa_{\s}c.
\end{align}
Introducing now two BRST-doublets $\cbh=\bar{c}$ and $\hat{B}=B$ with
\begin{alignat}{2}
\hat{s}\cbh &=  s\bar{c} & &= \hat{B} = B, \nn  \\
\hat{s}\hat{B} &= sB & &= 0 {   }, \label{hsB}
\end{alignat}
one has two possibilities to construct Faddeev-Popov actions:
\begin{equation}
\S^{(i)}_{\rm gf} = \int d^4x\,s \left(  \bar{c} \,\pa^{\m} A_{\m} + \frac{\a}{2} \bar{c}  B \right) = \int d^4x\,\left( B \,\pa^{\m} A_{\m} - \bar{c} \,\pa^2 c + \frac{\a}{2} B^2\right), \label{1}
\end{equation}
and
\begin{equation}
\hat{\S}^{(ii)}_{\rm gf} = \int d^4x\,\hat{s}\left( \cbh \,\pa^{\m} \hat{A}_{\m} + \frac{\a}{2} \cbh \hat{B}\right) = \int d^4x\,\left( B \,\pa^{\m} \hat{A}_{\m} - \bar{c} \,\pa^{\m} \hat{s}\hat{A}_{\m} + \frac{\a}{2} B^2 \right).
\end{equation}
An expansion in $\q$ of $\hat{\S}^{(ii)}_{\rm gf}$ leads to
\begin{multline}
{\S}^{(ii)}_{\rm gf}= \int d^4x\,\left\{ B \,\pa^{\m} A_{\m} - \bar{c} \,\pa^2 c+ \frac{\a}{2} B^2 \phantom{- \frac{1}{2}} \right. \\
\left. -\frac{1}{2}\,\q^{\r\s}\, [ B\,\pa^{\m} \left( A_{\r} \left( \pa_{\s}A_{\m} + F_{\s\m} \right) \right) + \bar{c}\,\pa^2 \left( \pa_{\r}c\,A_{\s} \right)  -2\,\bar{c}\,\pa^{\m} \left( \pa_{\r}c\,\pa_{\s}A_{\m} \right) ] \right\}. \label{2}
\end{multline}
The case described by (\ref{1}) is the usual linear $\pa^{\m}A_{\m}=0$ gauge, whereas the Faddeev-Popov action of (\ref{2}) corresponds to a highly nonlinear gauge. However, one has to stress that both cases lead to the same photon propagator
\begin{align}
{\langle A_{\m}A_{\n} \rangle }_{(0)} = \frac{1}{k^{2}+\im\e} \left( g_{\m\n}- (1-\a) \frac{k_{\m}k_{\n}}{k^{2}} \right) .
\end{align}
In terms of (\ref{sA}), (\ref{sc}) and (\ref{hsB}) the symmetry content of (\ref{1}) and (\ref{2}) may be expressed by a linearized Slavnov-Taylor identity
\begin{align}
\cs\,\S_{\q}^{(0)} = \int d^{4}x \left[ \pa_{\m}c\,\frac{\d\S_{\q}^{(0)}}{\d A_{\m}} + B\,\frac{\d\S_{\q}^{(0)}}{\d \bar{c}} \right] = 0, \label{S}
\end{align}
with
\begin{align}
\S_{\q}^{(0)} = \S_{\rm inv} + \left\{ \begin{array}{c} \S^{(i)}_{\rm gf} \\ \S^{(ii)}_{\rm gf} \end{array} \right. .\label{ta}
\end{align}
Formula (\ref{S}) implies the following transversality condition for the $2$-point $1$PI vertex functional
\begin{align}
\pa_{\m}^x\frac{\d^2\S_{\q}^{(0)}}{\d A_{\m}(x)\,\d A_{\n}(y)} = 0,
\end{align}
meaning that the ``vacuum polarization'' of the photon is transverse in momentum space
\begin{align}
p^{\m}\P_{\m\n}(p,-p) = 0. \label{trans}
\end{align}
In a companion paper \cite{Bichl:2001nf} we have studied perturbatively this transversality condition.

However, at this stage one has to state that the total action (\ref{ta}) is not the whole story. Because of the deformation parameter $\q^{\m\n}$ (which is treated as a constant external field) one can construct further gauge invariant quantities. Such additional terms in the BRST-invariant total action are in fact needed for the one-loop renormalization \cite{Bichl:2001nf} procedure.
\section{Higher Derivative Gauge Invariant Terms in the Action}
The presence of a constant external field $\q$ allows in principle to add ``infinitely'' many gauge invariant terms to the Lagrangian. Due to this fact one can construct the following bilinear BRST-invariant action in the photon sector\footnote{Note that $\int d^4x (- \frac{1}{2}\,F_{\m\n}\,\pa^2\ti{F}^{\m\n})$ and $\int d^4x (- \frac{1}{2}\,\pa^{\m}F_{\m\n}\,\pa^{\r}\ti{F}_{\r}^{\;\;\n})$ are of topological nature. Observe also that $\ti{\pa}A$ is gauge invariant.}
\begin{multline} \label{max}
\S_{\rm h.d.}^{\rm Max} = \int d^4x \left\{ - \frac{1}{2}\,F_{\m\n}\,\pa^2       \ti{F}^{\m\n} - \frac{1}{2}\,\pa^{\m}F_{\m\n}\,\pa^{\r}\ti{F}_{\r}^{\;\;\n}      - \frac{1}{4}\,F_{\m\n}\,\pa^2\ti{\pa}^2F^{\m\n} - \frac{1}{4}\,\q^2\,           F_{\m\n}\,(\pa^2)^2F^{\m\n} \right.   \\
\left. {} + \frac{1}{2}\,\left( \ti{\pa}^{\m}\pa^{\n}F_{\m\n} \right)^2 +        \frac{1}{2}\,\ti{F}_{\m\n}\,(\pa^2)^2\ti{F}^{\m\n}  +                            \mbox{higher derivative terms} \right\}, 
\end{multline}
with
\begin{align}
\ti{\pa}^{\m} &= \q^{\m}_{\;\;\n}\pa^{\n}, \nn 
\\
\dtt^{\m} &= \q^{\m}_{\;\;\r}\q^{\r}_{\;\;\s}\pa^{\s}, \nn
\\
\q^2 &= \q_{\m}^{\;\;\n}\q_{\n}^{\;\;\m}, \nn
\\
\ti{F}^{\m\n} &= \q^{\m\r}F_{\r}^{\;\;\n}, \nn
\\
\ti{F} &= \ti{F}_{\m}^{\;\;\m}. \label{thetadef}
\end{align}
Alternatively, (\ref{max}) may be rewritten as
\begin{multline}
\S_{\rm h.d.}^{\rm Max} = \int d^4x \left\{ - \frac{1}{2}\,F_{\m\n}\,\pa^2       \ti{F}^{\m\n} - \frac{1}{2}\,\pa^{\m}F_{\m\n}\,\pa^{\r}\ti{F}_{\r}^{\;\;\n} -    \frac{1}{4}\,F_{\m\n}\,\pa^2\ti{\pa}^2F^{\m\n} \right. \\
 - \frac{1}{2}\,A^{\m} \left[ \ti{\pa}^2 (\pa^2)^2 g_{\m\n} + (\pa_{\m}\dtt_{\n} + \pa_{\n}\dtt_{\m})(\pa^2)^2 
+ (\pa^2)^3\,\q^{\r}_{\;\;\m}\,\q_{\r\n} \right] A^{\n} \\
\left.  - \frac{1}{4}\,\q^2\,F_{\m\n}\,(\pa^2)^2F^{\m\n} 
 + \frac{1}{2}\,\ti{\pa}A\,(\pa^2)^2\ti{\pa}A + \mbox{higher derivative terms} \right\}.
\end{multline}
In a similar way one can construct with $\ti{D}_{\m}=\q_{\m}^{\;\;\n}D_{\n}\,\,\,(D_{\m}=\pa_{\m}-\im A_{\m})$ also higher derivative terms for the Dirac theory
\begin{multline} \label{dir}
\S_{\rm h.d.}^{\rm Dir} = \int d^4x \left\{ m\,\bar{\j} \left( D\ti{D} +         (D\ti{D})^2 +  D^2\ti{D}^2 + \ldots \right) \j \right. \\
+ \bar{\j}\g^{\m}\,\im D_{\m} \left( D\ti{D} + (D\ti{D})^2 +  D^2\ti{D}^2 +      \ldots \right) \j  \\
\left.  + \bar{\j}\g^{\m}\,\im \ti{D}_{\m}\,D^2\left( 1 + D\ti{D} + (D\ti{D})^2  +  D^2\ti{D}^2 + \ldots \right) \j + \mbox{h.d.} \right\}. 
\end{multline}
It is interesting to notice that one can construct also four-fermion vertices. For example one has
\begin{align}
\S_{\rm (4),h.d.}^{\rm Dir} = \int d^4x \left\{ \bar{\j}\ti{D}_{\m}\j\bar{\j}\ti{D}^{\m}\j + \bar{\j}\ti{D}_{\m} \left( D\ti{D} + (D\ti{D})^2 +  D^2\ti{D}^2 + \ldots \right)\j\bar{\j}\ti{D}^{\m}\j + \mbox{h.d.} \right\}. \label{4dir}
\end{align}
However, these terms cannot be obtained from a truly noncommutative action by the Seiberg-Witten map because the corresponding expression would be non-local. Therefore, we expect that contributions of this type, which are possible for individual Feynman graphs, cancel on the level of Green's functions. 
Equations (\ref{max})--(\ref{4dir}) imply that even at the classical level one has $\q$-dependent terms. This entails the following form of the ``vacuum polarization'' of the photon in momentum space representation
\begin{equation}
  \begin{split}
    {\P}^{(0)}_{\m\n}(p,-p; \q) \approx (p^{2})^2\q^2 \left(p_{\m}p_{\n} - g_{\m\n}p^{2} \right)+\ti{p}^{2}p^{2} \left(p_{\m}p_{\n} - g_{\m\n}p^{2} \right)+(p^{2})^2\ti{p}_{\m}\ti{p}_{\n}   \\
  +(p^{2})^2\left( g_{\m\n}\ti{p}^{2}+p_{\m}\Tilde{\Tilde{p}}_{\n}+p_{\n}           \Tilde{\Tilde{p}}_{\m}+ p^{2}\q^{\r}_{\;\;\m}\q_{\r\n}\right),                   \label{transx}
  \end{split}
\end{equation}
in observing that no terms linear in $\q^{\m\n}$ occur.
Due to
\begin{align}
p\ti{p} = 0, \qquad p\Tilde{\Tilde{p}} = -\tilde{p}^2,
\end{align}
it is obvious that (\ref{transx}) fulfills the transversality condition (\ref{trans}), where we have used a notation analogous to (\ref{thetadef}). In order to be complete the total action is given by
\begin{align}
\S_{\q}^{(0)} = \S_{\rm inv} + \left\{ \begin{array}{c} \S^{(i)}_{\rm gf} \\ \S^{(ii)}_{\rm gf} \end{array} \right\} + \S_{\rm h.d.}^{\rm Max} + \S_{\rm h.d.}^{\rm Dir} + \S_{\rm (4),h.d.}^{\rm Dir}.
\end{align}
In the limit $\q \rightarrow 0$ one recovers the ordinary QED with a free Faddeev-Popov sector. One has to stress also that equation (\ref{max}) is required for the renormalization procedure at the one-loop level for the calculation of ${\P}^{(1)}_{\m\n}(p,-p;\q)$ \cite{Bichl:2001nf}.
\section{Non-Abelian Extension: Pure Yang-Mills Case (NCYM)}
We start with the non-Abelian extension of (\ref{SWQED}) in considering only the gluon sector. The corresponding gluon field is now Lie-algebra valued:
\begin{align}
A_{\m}(x) = A_{\m}^a(x)T^a,
\end{align}
where the $T^a$ are the generators of a $U(N)$ gauge group with
\begin{gather}
[T^a,T^b] = \im f^{abc}T^c, \quad \{T^a,T^b\} = d^{abc}T^c, \nn \\
\tra\left( T^aT^b \right) = \d^{ab}, \quad T^0 = \frac{1}{\sqrt{N}}\,\mathbbm{1}.
\end{gather}
The corresponding non-Abelian field strength is defined as usual
\begin{align}
F_{\m\n} = \pa_{\m}A_{\n}-\pa_{\n}A_{\m}-\im \left[ A_{\m},A_{\n} \right] .
\end{align}
The infinitesimal gauge transformations are
\begin{align}
\d_{\l}A_{\m} &= \pa_{\m}\l + \im \left[ \l,A_{\m} \right] \equiv D_{\m}\l, \nn
\\
\d_{\l}F_{\m\n} &= \im \left[ \l,F_{\m\n} \right] . \label{ini}
\end{align}
As shown in \cite{Seiberg:1999vs} the Seiberg-Witten map for the non-Abelian extension is
\begin{align}
\hat{A}_{\m} &= A_{\m} + A'_{\m} = A_{\m} -\frac{1}{4}\,\q^{\r\s} \left\{ A_{\r},\pa_{\s}A_{\m} + F_{\s\m} \right\} + O(\q^2), \label{swA}
\\
\hat{\l} &= \l + \l' = \l + \frac{1}{4}\,\q^{\r\s} \left\{ \pa_{\r}\l,A_{\s} \right\} + O(\q^2). \label{swl}
\end{align}
From
\begin{align}
\hat{F}_{\m\n} = \pa_{\m}\hat{A}_{\n}-\pa_{\n}\hat{A}_{\m}-\im \left[ \hat{A}_{\m},\hat{A}_{\n} \right]_{\star} 
\end{align}
follows with the help of (\ref{swA})
\begin{align}
\hat{F}_{\m\n} = F_{\m\n} + \frac{1}{4}\,\q^{\r\s} \left( 2 \left\{ F_{\m\r},F_{\n\s} \right\} - \left\{ A_{\r},D_{\s}F_{\m\n} + \pa_{\s}F_{\m\n} \right\} \right) + O(\q^2). \label{f}
\end{align}
Corresponding to (\ref{ini}) one has 
\begin{align}
\hat{\d}_{\hat{{\l}}}\hat{A}_{\m} &= \pa_{\m}\hat{\l} + \im \left[ \hat{\l},\hat{A}_{\m} \right]_{\star}, \nn
\\
\hat{\d}_{\hat{{\l}}}\hat{F}_{\m\n} &= \im \left[ \hat{\l},\hat{F}_{\m\n} \right]_{\star} . \label{sw}
\end{align}
In the sense of Seiberg and Witten (\ref{sw}) implies
\begin{equation}
\hat{\d}_{\hat{{\l}}}\hat{A}_{\m} = \pa_{\m}\hat{\l} + \im \left[ \hat{\l},\hat{A}_{\m} \right] -\frac{1}{2}\,\q^{\r\s} \left( \pa_{\r}\l\,\pa_{\s}A_{\m} - \pa_{\r}A_{\m}\,\pa_{\s}\l \right) + O(\q^2).
\end{equation}
Inserting (\ref{swA}) and (\ref{swl}) one arrives finally at
\begin{multline}
\hat{\d}_{\hat{{\l}}}\hat{A}_{\m} = D_{\m}\l + \frac{1}{4} \,\q^{\r\s} \left( \left\{ \pa_{\m}\pa_{\r}\l,A_{\s} \right\} + \left\{ \pa_{\r}\l,\pa_{\m}A_{\s} \right\} \right)   \\
 +  \frac{\im}{4}\,\q^{\r\s} \left( \left[ \left\{ \pa_{\r}\l,A_{\s} \right\},A_{\m} \right] - \left[ \l, \left\{ A_{\r},\pa_{\s}A_{\m}+F_{\s\m} \right\} \right] \right) + O(\q^2).
\end{multline}
Now we are ready to construct the gauge invariant non-Abelian action. Following \cite{Jurco:2000ja} one has at the classical level
\begin{align}
\hat{\S}_{\rm inv} =  - \frac{1}{4g^2} \int d^4x\, \tra\left(\hat{F}_{\m\n}\star\hat{F}^{\m\n}\right) =  - \frac{1}{4g^2} \int d^4x\, \tra\left( \hat{F}_{\m\n}\hat{F}^{\m\n} \right). \label{action}
\end{align}
Using (\ref{f}) one gets for (\ref{action})
\begin{equation}  \label{hint}
  \begin{split}
\S_{\rm inv} &= - \frac{1}{4g^2} \int d^4x\, \tra\left( F_{\m\n}F^{\m\n} + \frac{1}{2}\,\q^{\r\s} \left( 2 \left\{ F_{\m\r},F_{\n\s} \right\}F^{\m\n} - \left\{ A_{\r},D_{\s}F_{\m\n} + \pa_{\s}F_{\m\n} \right\}F^{\m\n} \right)\right)   \\
&=  - \frac{1}{4g^2} \int d^4x\, \left( F^a_{\m\n}F^{a\,\m\n} \phantom{- \frac{1}{2}}\right.\\ 
& \left. \phantom{{=} - \frac{1}{4g^2} \int d^4x\,} + \q^{\r\s} \left( F^a_{\m\r}F^b_{\n\s}F^{c\,\m\n} - \frac{1}{2}\,A^a_{\r}(D_{\s}F_{\m\n})^bF^{c\,\m\n} - \frac{1}{2}\,A_{\r}^a\,\pa_{\s}F^b_{\m\n}F^{c\,\m\n} \right) d^{abc} \right). 
  \end{split}
\end{equation}
\section{Gauge-Fixing for Pure Yang-Mills Theory}
Similar to the QED case there are again two possibilities of gauge-fixing for the non-Abelian generalization. One replaces the infinitesimal gauge parameters by the corresponding ghost fields:
\begin{equation}
\l\rightarrow c, \quad \hat{\l}\rightarrow \hat{c}. \label{poss}
\end{equation}
The first choice leads to
\begin{align}
sA_{\m} &= \pa_{\m}c + \im \left[ c,A_{\m} \right] = D_{\m}c, \nn
\\
sc &= \im\,c^2 = \frac{\im}{2} \left\{ c,c \right\}. \label{sch} 
\end{align}
With the second possibility of (\ref{poss}) one gets from (\ref{swl})
\begin{align}
\hat{c} &= c + \frac{1}{4}\,\q^{\r\s} \left\{ \pa_{\r}c,A_{\s} \right\} + O(\q^2). \label{c}
\end{align}
Equations (\ref{c}) and (\ref{swA}) together with (\ref{sch}) allow now to calculate the BRST-variations for $\hat{A}_{\m}$ and $\hat{c}$:
\begin{align}
\hat{s}\hat{A}_{\m} &= sA_{\m} -\frac{1}{4}\,\q^{\r\s} \left( \left\{ sA_{\r},\pa_{\s}A_{\m} + F_{\s\m} \right\} + \left\{ A_{\r},\pa_{\s}sA_{\m} + sF_{\s\m} \right\} \right) , \nn
\\
\hat{s}\hat{c} &= sc + \frac{1}{4}\,\q^{\r\s} \left( \left\{ \pa_{\r}sc,A_{\s} \right\} - \left[ \pa_{\r}c,sA_{\s} \right] \right) . \label{h}
\end{align}
By explicit calculations one verifies that the equations (\ref{h}) follow by $\q$-expansion from
\begin{align}
\hat{s}\hat{A}_{\m} &= \pa_{\m}\hat{c} + \im \left[ \hat{c},\hat{A}_{\m} \right]_{\star} , \nn
\\
\hat{s}\hat{c} &= \im\,\hat{c}\star\hat{c}.
\end{align}
The above results may be summarized by the Seiberg-Witten map for the BRST-transformation 
\begin{align}
\hat{A}_{\m} \left( A_{\n} \right) + \hat{s}\hat{A}_{\m}(A_{\n}) &= \hat{A}_{\m} \left( A_{\n} + sA_{\n} \right).
\end{align}

For practical reasons we choose now the simpler possibility for the gauge-fixing procedure:
\begin{align}
\S_{\rm gf} &= \int d^4x\,\tra \left(s( \bar{c} \,\pa^{\m} A_{\m} + \frac{\a}{2} \bar{c} B )  \right) = \int d^4x\tra\left( B \,\pa^{\m} A_{\m} - \bar{c} \,\pa^{\m} \left( \pa_{\m}c + \im \left[ c,A_{\m} \right] \right)+\frac{\a}{2}B^2 \right),
\end{align}
where $\bar{c}$ and $B$ are the corresponding non-Abelian antighost and multiplier field.
To be complete one also adds further gauge invariant pieces to the total gauge-fixed action. They are obtained from (\ref{max}) by replacing the derivatives by the covariant derivatives and the Abelian by the non-Abelian field strength. The total BRST-invariant action is thus
\begin{multline} \label{arni}
\S_{\q}^{(0)} = {\S}_{\rm inv} + \S_{\rm gf} + \S_{\rm h.d.} 
\\
= - \frac{1}{4g^2}\int d^4x\, \tra\left( F_{\m\n}F^{\m\n} + \frac{1}{2}\,\q^{\r\s} \left( 2 \left\{ F_{\m\r},F_{\n\s} \right\}F^{\m\n} - \left\{ A_{\r},D_{\s}F_{\m\n} + \pa_{\s}F_{\m\n} \right\}F^{\m\n} \right)  \right) 
\\
+ \int d^4x\, \tra\left( B \,\pa^{\m} A_{\m} - \bar{c} \,\pa^{\m} \left( \pa_{\m}c + \im \left[ c,A_{\m} \right] \right) + \frac{\a}{2} B^2 - \frac{1}{2}\,F_{\m\n}\,D^2\ti{F}^{\m\n} \right.
\\
 - \frac{1}{2}\,D^{\m}F_{\m\n}\,D^{\r}\ti{F}_{\r}^{\;\;\n} - \frac{1}{4}\,F_{\m\n}\,D\ti{D}F^{\m\n} - \frac{1}{4}\,F_{\m\n}\,D^2\ti{D}^2F^{\m\n} - \frac{1}{4}\,\q^2\,F_{\m\n}\,(D^2)^2F^{\m\n} 
\\
\left. + \frac{1}{2}\,\left( \ti{D}^{\m}D^{\n}F_{\m\n} \right)^2 + \frac{1}{2}\,\ti{F}_{\m\n}\,(D^2)^2\ti{F}^{\m\n} + \mbox{higher derivative terms} \right). 
\end{multline}
In order to characterize the BRST-symmetry  of this model one has to introduce BRST-invariant external sources $\r^{\m}, \s$ allowing to describe the non-linear BRST-transformations consistently. This leads to
\begin{equation}
\S_{\rm ext} = \int d^4x\, \tra \left( \r^{\m}sA_{\m} + \s sc \right)
\end{equation}
and to the action 
\begin{equation}
\S_{\q,\rm{tot}}^{(0)} = \S_{\q}^{(0)} + \S_{\rm ext}.
\end{equation}
The BRST-symmetry is therefore encoded at the classical level by the non-linear Slavnov-Taylor identity
\begin{equation}
\cs\,\S_{\q,\rm{tot}}^{(0)} = \int d^{4}x \tra \left[ \frac{\d\S_{\q,\rm{tot}}^{(0)}}{\d \r^{\m}}\,\frac{\d\S_{\q,\rm{tot}}^{(0)}}{\d A_{\m}} + \frac{\d\S_{\q,\rm{tot}}^{(0)}}{\d \s}\,\frac{\d\S_{\q,\rm{tot}}^{(0)}}{\d c} + B\,\frac{\d\S_{\q,\rm{tot}}^{(0)}}{\d \bar{c}} \right] = 0.
\end{equation}
\section{Conclusion and Outlook}

We have defined a $\q$-deformed QED with the help of Seiberg-Witten maps for photon and fermion fields at the classical level. In the same manner we also discussed a $\q$-deformed Yang-Mills model.

In a next step we focus on the quantization procedure. Some preliminary work has been  done in \cite{Bichl:2001nf}, where the quantization of a $\q$-deformed Maxwell theory is presented. There, one-loop corrections to the 1PI two-point function for the photon are investigated in the presence of non-renormalizable interaction vertices.

\section{Acknowledgement}
We would like to thank Harald Grosse, Karl Landsteiner, Stefan Schraml, Raymond Stora, and Julius Wess for enlightening discussions.

\end{document}